# The Nuclear Liquid-Gas Phase Transition: Q.E.D

V.E. Viola[1*]

Department of Chemistry and IUCF
Indiana University, Bloomington, IN 47405

For the past decade, intense experimental effort has been devoted to the search for a liquid-gas phase transition in highly excited nuclei. Now, synthesis of the large body of existing multifragmentation data provides a strong case for identification of this phenomenon. In this presentation we discuss several salient features of the data that support their interpretation in terms of a spinodal liquid-gas phase transition.

## 1. INTRODUCTION

The nuclear liquid-gas phase transition [1,2] touches on several challenging physics problems. From the experimental point of view, the goal has been to demonstrate the link (or lack thereof) between multifragmentation events and the general properties expected for a phase transition – and eventually to extend these studies to highly neutron/proton asymmetric matter. For theorists, there is the need to describe the dynamical evolution of these complex reactions and to extract information relevant to the nuclear equation of state. Finally, the question arises for the astrophysicists: what can these results tell us about the conditions in the core of a collapsing supernova? .

In this paper we focus on the numerous experimental signals for spinodal decomposition that give rise to interpretation of the data in terms of a nuclear liquid-gas phase transition and perhaps, critical behavior. These signals have been produced using collisions between heavy nuclei and an array of projectiles, ranging from GeV hadrons to gold ions, at several accelerators (AGS, Dubna, GANIL, GSI, LBL, LNS Saclay, NSCL and Texas A&M). Many of the examples presented here are from light-ion-induced reactions obtained by the ISiS collaboration. But regardless of the method of preparation, the multifragmentation events are remarkably similar [3], as is discussed in several related papers at this meeting.

## 2. COLLISION DYNAMICS

Two general approaches have been employed to prepare the hot nuclear systems that are candidates for observing a nuclear liquid-gas phase transition: symmetric A + A reactions in the Fermi-energy regime (E/A ~ 50 MeV/nucleon) and very asymmetric hadron(h) + A reactons at bombarding energies of several GeV. Heating via the A + A path introduces important variables such as compression, angular momentum and the physics-rich neck region at the interface between the two colliding ions. In contrast, h +

[1] For the ISiS LNS Saclay E228 and Brookhaven AGS E900 Collaborations
[*] Research support from the U.S. Department of Energy Grant No. DE.FG02-88ER.40404A019

A reactions minimize these effects, creating only a single source and isolating the thermal aspects of the problem.

Simulations based on variations of the Boltzmann equation [4 and refs therein] have provided valuable insights into the time evolution of the collision process. The paper by B. Borderie in these proceedings describes a current A + A simulation. For h+A systems, Turbide *et al.* [5] have recently employed the BUU simulation of Danielewicz [6] to examine predictions for 6-15 GeV/c p + $^{197}$Au reactions. This code includes in-medium cross sections, a momentum-dependent potential and the formation of A = 2 and 3 clusters. The calculations indicate that the hot residues are randomized (constant entropy/nucleon) after a time of $\tau \sim 30$ fm/c, and for the most central impact parameters, are produced in a state of high excitation energy E* and depleted density as a result of the fast cascade process. Comparison with experimental E* distributions and p/d/t/$^3$He ratios predicts that fast cascade/preequilibrium emission continues for another ~30 fm/c, after which the residues are essentially thermalized; i.e., equilibrium-like behavior appears to be achieved after about 60-70 fm/c.

## 3. EXPERIMENTAL SIGNALS

Of the many experimental signals for a phase transition in hot nuclei, perhaps the most transparent are found in the evolution of the spectra as a function of excitation energy (E*/A). In Fig. 1 invariant cross sections ($v_\parallel$ vs. $v_\perp$) are shown for hydrogen and carbon ions emitted in the 8.0 GeV/c $\pi^-$ + $^{197}$Au reaction for three different excitation energies. For energetic protons ($v \gtrsim 0.2$ c) there is a spray of forward-emitted particles that originate in fast-cascade/preequilibrium processes. This component gradually diminishes with Z, so that for carbon and heavier ions it is of minor importance. The low-energy component of these spectra (radial cuts), however, is nearly symmetric about zero velocity, indicating emission from a randomized (equilibrated?) source that is moving with a velocity of ~ 0.01c. This thermal-like component accounts for up to 5% of the total cross section, or about 100 mb.

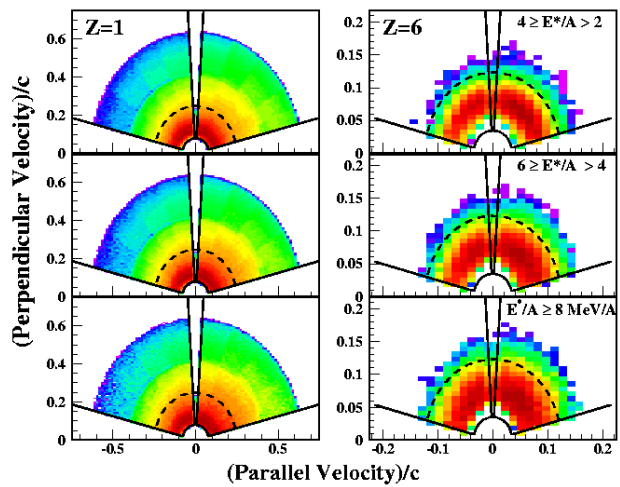

Figure 1. Contour plot of longitudinal v vs. transverse v velocity of hydrogen (left) and carbon (right) fragments from the 8.0-GeV/c $\pi^-$ + $^{197}$Au reaction for several bins in E*/A. Solid lines indicate geometrical acceptance of the ISiS array; dashed line gives the thermal cutoff velocity [11], not corrected for source velocity.

Separating the equilibrium events from those having a nonequilibrium origin (outside the dotted lines in Fig. 1) must be done empirically from analysis of the spectra [7].

The evolution of the spectra with E*/A yields further insight into the multifragmentation mechanism. Of primary importance is the behavior of the Coulomb-like peaks for IMFs (IMF: $3 \lesssim Z \lesssim 15$), which systematically decrease in energy with increasing E*/A, instead of increasing, which would be expected for increasing thermal energy. This behavior can be explained in terms of emission from an expanded/dilute source, another predicted property for the phase transition. This behavior is investigated further in Sec. 4. In contrast, the IMF spectral slope temperature parameters increase systematically with E*/A [8]. The net result of the reduced Coulomb peaks and increased slope-temperature parameters is an average kinetic energy as a function of Z that is nearly independent of excitation energy [9].

While the data shown in Fig. 1 support a statistical interpretation of the thermal (low-energy) component of the multifragmentation cross section, statistical behavior alone does not distinguish between sequential evaporative emission and an instantaneous spinodal mechanism. Nor does it give insight into possible critical phenomena. For this purpose it is essential to know the breakup time scale, which can be extracted from measurements of IMF-IMF correlation functions [10]. The results of such an analysis for the 8.0 GeV $\pi^-$ + $^{197}$Au reaction are shown as a function of E*/A in the bottom panel of Fig. 2. At excitation energies near E*A ~ 2 MeV the time scale is found to be of order 500 fm/c, consistent with expectations for evaporative emission. Above this value the time scale decreases rapidly, reaching 20-50 fm/c for E*/A ~ 5 MeV and above. Such a time scale implies a nearly spontaneous disintegration process. Thus, the excitation energy region between E*/A ~ 2 – 5 MeV appears to involve rapid changes in the emission time and a transition from sequential to instantaneous decomposition.

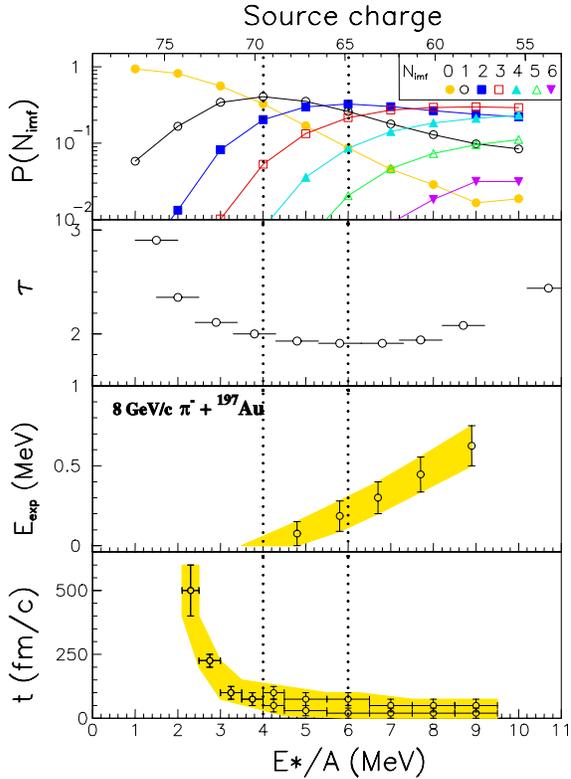

Figure 2. Dependence on E*/A for the following quantities, from bottom up: relative IMF emission time τ, extra radial expansion energy $E_{exp}/A_{IMF}$, charge distribution power exponent $Z^{-\tau}$, and probability for a given IMF multiplicity $P(N_{IMF})$.

Accompanying the rapid decrease in emission time in the E*/A = 2-5 MeV range are distinct changes in the character of many reaction variables. In the top frame of Fig. 2, the probability of emitting IMFs with a given multiplicity $N_{IMF}$ is plotted as a function of E*/A. Below E*/A ~ 4 MeV multiple IMF events are rare. But with increasing excitation energy, multiple IMF emission begins to dominate the cross section and the number of IMFs per event, along with light-charged particles and neutrons, grows systematically. Above E*/A ~ 5 MeV, only IMFs and light particles are observed [11].

The second panel from the top in Fig. 2 shows the dependence of the fragment size (Z) distribution as a function of excitation energy, expressed in terms of the power law exponent tau derived from the fragment charge distributions, $\sigma(Z) \propto Z^{-\tau}$. In the interval E* ≃ 4-6 MeV the power law exponent reaches a minimum, indicating that IMFs of maximum size are formed in this region of excitation, as expected for spinodal breakup of the system (see also Sec. 5). Another correlation shown in the next-to-bottom panel of Fig. 2 is the extra radial expansion energy; i.e., the expansion energy in excess of that required to reach the breakup configuration. While the observed values are much smaller than found in heavy-ion reactions, where compression-decompression effects are present, extra expansion energy due to thermal effects appears to play a finite role above E*/A ~ 5 MeV. Finally, as discussed in the Sec. 4, the breakup density decreases rapidly between E*/A ≃ 2 and 5 MeV.

In summary, numerous experimental observables all point to a dramatic change in the reaction mechanism in the interval E*/A ~ 2-5 MeV – with properties that are consistent with a first- or second-order liquid-gas phase transition.

## 4. THERMODYNAMICS: THE CALORIC CURVE AND HEAT CAPACITY

One of the most stimulating early interpretations of multifragmentation data was the heat versus temperature, or "caloric curve", proposed by the GSI group [12]. By plotting temperatures derived from double-isotope ratios [13] as a function of excitation energy, a temperature versus heating curve was obtained that is reminiscent of that for heating liquid water to the boiling point. Subsequent experiments generated numerous caloric curves, not all of which were in agreement with one another. Recently Natowitz *et al.* have shown that by analyzing the existing data as a function of the mass of the fragmentating source, a systematic understanding of the data is obtained [14]. In Fig. 3 the temperatures corresponding to the plateau regions decrease with increasing source mass, an effect that is attributed to the source temperature having reached the Coulomb instability limit [15]. From these systematics it was possible to derive a value of the critical temperature for infinite nuclear matter of 16 ± 1 MeV and a nuclear compressibility constant of K = 232 ± 30 MeV [16] in good agreement with the Giant Monopole Resonance value of K = 230 ± 5 MeV [17].

One major uncertainty in the caloric curve result has been the use of temperatures derived from double isotope ratios, which are sensitive to detector acceptance and preequilibrium contamination of the spectra. An alternative approach suggested by Natowitz *et al.* [18] is to correlate the breakup density with temperature via a density-dependent Fermi gas model,

$$T = \sqrt{(\rho/\rho_0)^{2/3} E*/a} \quad , \tag{1}$$

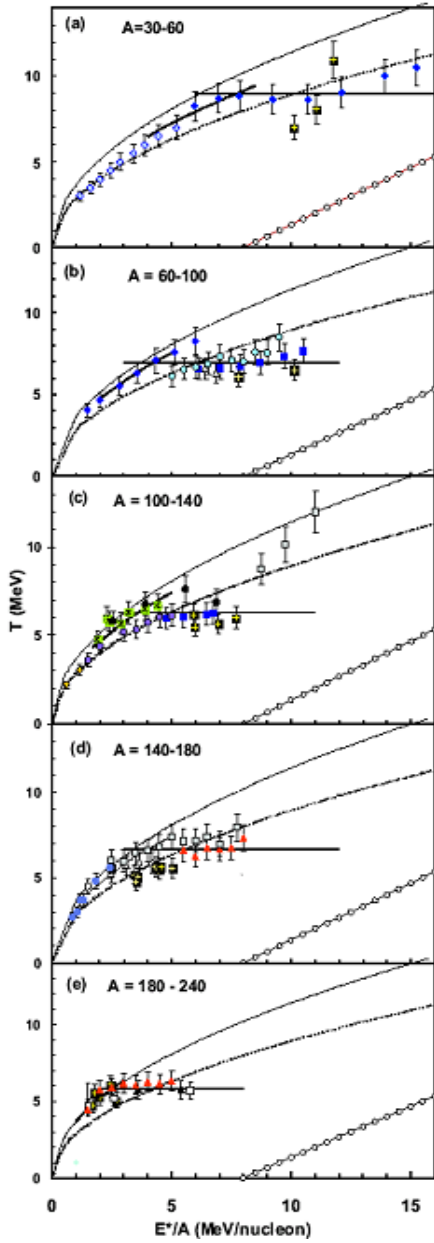

Figure 3. Caloric curves for five selected regions of mass from Natowitz *et al.*, [14]. Solid horizontal lines are interpreted as the temperature at which the Coulomb instability limit is reached. Broken lines are predictions described in [14].

where $\rho/\rho_0$ is the ratio of the breakup density relative to normal nuclear matter density, *a* is the level-density parameter, and the effective mass of the nucleon m* is assumed to be unity. Recently, breakup densities have been derived from moving-source analyses of the spectral peaks of IMFs measured in reactions of several projectiles with $^{197}$Au nuclei covering a wide range of excitation energies [19]. As shown in the left frame of Fig. 4, the breakup density decreases systematically from $\rho \simeq \rho_0 \sim 1$ at E* ~ 2 MeV and below, to $\rho/\rho_0$ ~0.25-0.30 at E*/A ~ 5 MeV and above – consistent with the observables shown in Fig. 2 and the caloric curve behavior in Fig. 3. By using the measured densities and the Fermi-gas relationship in Eq. (1), a caloric curve can be generated that is *independent of any other temperature assumptions*. The density-dependent Fermi gas caloric curve is shown in the right frame of Fig. 4 and is seen to be in good agreement with those obtained using double-isotope ratio temperatures in Fig. 3 for this mass region. Also shown in Fig. 4 is the expected Fermi gas prediction for normal nuclear density, using a value of a = A/11.3 MeV$^{-1}$ obtained from an empirical fit to the E*/A data below 2 MeV/nucleon. The implication of this analysis is that caloric curve behavior is driven by thermal expansion/dilution and that the Coulomb instability limit is reached near T ~ 5 MeV and E*/A ~ 4 MeV for this A ~ 150-200 system.

Another thermodynamic signal of the phase transition is the appearance of a negative heat capacity at the transition point. D'Agostino *et al.* have demonstrated this effect from analysis of the fluctuations in the fragment kinetic energies [20]. Their analysis reveals the presence of a negative branch in the heat capacity curve near E*/A ~ 5 MeV for an A ~ 200 system, in line with all the earlier signals. An alternative approach used by Das *et al.* [21] differentiates the caloric curve to obtain the heat capacity and also finds a negative excursion in the heat capacity near $E^*/A \sim 4-5$ MeV. The concordance that is now emerging from both the caloric curve and heat capacity analyses provides still another strong argument in favor of a liquid-gas phase transition, suggestive of a first-order phase transition.

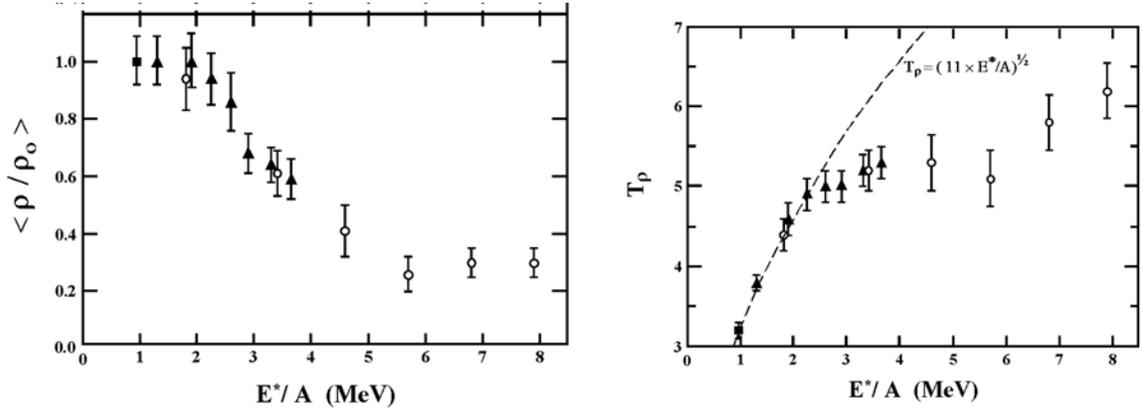

Figure 4. Left: Average source density $<\rho/\rho_0>$ derived from moving-source-fit Coulomb parameters [19] as a function of excitation energy. Right: Caloric curve based on density-dependent Fermi-gas temperatures [18].

## 5. SCALING LAWS

An additional criterion that must be met in order to argue in favor of a liquid-gas phase transition and possibly critical behavior is the agreement of the data with various scaling laws. This type of analysis was pioneered by the Purdue group in the early 1980's and subsequently applied to the EOS $^{197}$Au + $^{12}$C studies [22]. One such test of statistical behavior is binomial scaling [23]. Multifragmentation multiplicity distributions from the 8.0 GeV/c $\pi^-$ + $^{197}$Au reaction can be described to a high degree of accuracy with a binomial scaling law, lending additional support to the statistical nature of the breakup process [24]. Recently two important scaling tests of ISiS data have been published, using the Fisher droplet model [25] and percolation theory [26]. In the left frame of Fig. 5 several hundred $Z \geq 5-15$ data points from the 8.0 GeV/c $\pi^-$ + $^{197}$Au reaction have been fit with a six-parameter Fisher droplet model [27]. The fitting parameters are found to be consistent with expectations for a nuclear liquid drop and the results are interpreted in terms of a first-order phase transition and critical behavior with critical temperature $T_c$ = 6.7 MeV for the finite system. The right-hand side of Fig. 5 shows a percolation analysis for Z = 3-7 fragments from the 10.2 GeV/c p + $^{197}$Au reaction [28]. This analysis concludes the phase transition is second order and finds a critical temperature of $T_c$ = 8.3 MeV for the finite system.

While uncertainties may exist with respect to the order of the phase transition and the value of $T_c$, in particular the treatment of Coulomb energy, one cannot escape the fact that the scaling laws are highly successful in describing the IMF distributions and thus provide another strong case for a phase transition and possibly critical behavior.

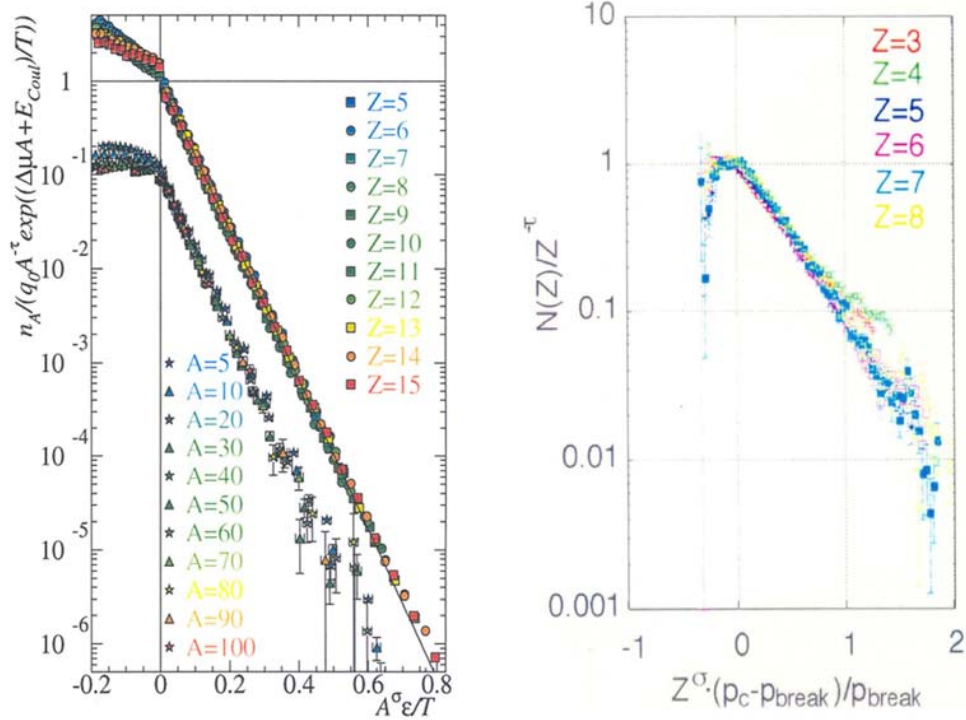

Figure 5. Left: Fisher droplet model fits to Z = 5-15 IMF data [27] from 8.0 GeV/c $\pi^-$ + $^{197}$Au reaction. Right: Percolation model fits to Z = 3-7 data, corrected for sequential decay [28], from 10.2 GeV/c p + $^{197}$Au reaction.

## 6. CONCLUSIONS

Given the wealth of multifragmentation data that now exist, one can conclude with confidence that we have observed the spinodal decomposition of a finite nuclear system. For reactions induced by GeV hadrons on $^{197}$Au, comparison of BUU simulations with the data indicate that the heavy residue is randomized/equilibrated after about 60 fm/c. All of the experimental observables – multiplicities, fragment distributions, time scale, source density, etc. – undergo a distinct, self-consistent change in character over the excitation energy range E*/A ~ 2-5 MeV. In this range, caloric curve behavior and negative heat capacity are observed. Above E*/A ~ 5 MeV, spontaneous breakup into clusters and nucleons occurs on a 20-50 fm/c time scale from a dilute source with density $\rho/\rho_0$ ~ 0.3.

Perhaps the most important message of this international effort is that spinodal decomposition of the nuclear liquid has been identified. The task that remains is to fine-tune the thermodynamic parameters of this process. On the horizon is dependence of these properties on neutron-proton asymmetry.

The author wishes to acknowledge his many collaborators who participated in ISiS experiments E228 at LNS Saclay and E900 at the Brookhaven AGS, and in particular his colleague Kris Kwiatkowski.